# The journey of the German expedition to observe the transit of Venus on December 9, 1874 at the Kerguelen Islands and its sojourn there

Emmanuel Davoust
*Observatoire Midi-Pyrénées*

**Abstract.** Astronomers are the surveyors of the sky. They have always dedicated much time and resources to determining the scale of distances in the universe. Today, the Hubble constant; yesterday, the solar parallar. The following account, attributed to Ladislas Weinek, a modest actor of this scientific epic, describes one of its numerous chapters.

**Résumé.** Les astronomes sont les arpenteurs du ciel. Ils ont toujours consacré beaucoup de temps à déterminer l'échelle des distances dans l'univers. Aujourd'hui la constante de Hubble, hier le parallaxe solaire. Le récit qui suit, attribué à Ladislas Weinek, un modeste acteur de cette épopée scientifique, est l'un de ses nombreux chapitres.

My name is Ladislas Weinek, and I have been a Professor of astronomy at the German University and Director of the Imperial Observatory for nearly 28 years. Now that the end is near, the leisure time accorded to me by these two duties is spent upon meditation on my past, while I stroll alone through the little streets of our old city of Prague.

If future generations remember me at all, it will probably be for my work on selenography. But the event that stands out in my career, and that I always remember with great nostalgia, is my journey to the antipodes, nearly 37 years ago, to observe the transit of Venus across the Sun. I have often relived this expedition in my mind, in an attempt to find and analyze all the sensations and emotions linked to each stage, occasionally leafing through my drawing pads of the time, and my memories remain as precise as though it had all happened last year.

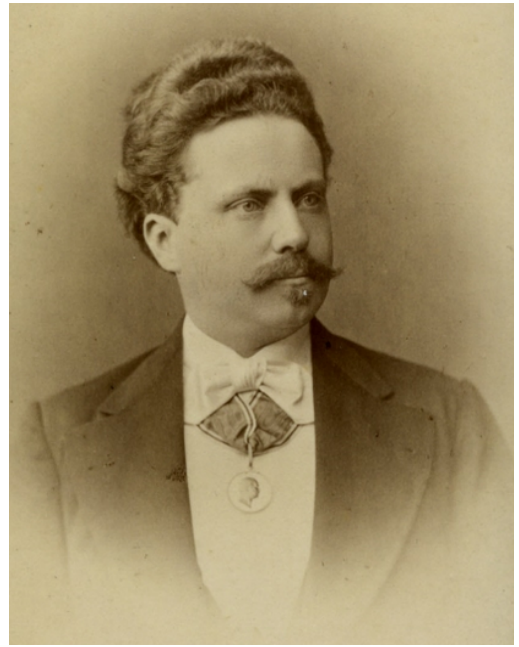

*Ladislas Weinek*[1]

### The project

It all began in 1869, when I was an observer at the Royal Saxon Observatory in Leipzig, after receiving my diploma as Candidate in Science at that city's University. When the English, French and Russians began to organize expeditions to observe the next transit of Venus, the director of our establishment, Professor Brunhs, was one of three German astronomers who were concerned that Germany would be left out of this scientific adventure, as it had been during the two previous transits, in 1761 and 1769. They thus formed a committee, which the great names of German astronomy soon joined. The stakes were considerable, because the distance from the Earth to the Sun would at last be precisely determined, and from there, the scale of distances in the solar system.

This committee, which convened in Berlin, was initially slowed down in its work by the war with France. But the proclamation of the German Empire in January, 1871, gave it an impetus, and subsequently the preparation was carried on with all the rigor and efficiency that characterize the Prussian mind, and with the stiffness and formality that are its inevitable counterpart. Each time that he was back in Leipzig, Professor Bruhns added to our information by conferences, along with personal remarks that provided insight into the scientific and political stakes of this project, as well as into the powerful personalities who conceived and organized it to the finest detail.

---
1  Image © Staatliche Museen zu Berlin, Kunstbibliothek – Sammlung Fotografie

The first problem addressed by the committee was the method of observation. The one that required measuring the two successive moments of contact of the planet with the two edges of the Sun, used in the 18th century, was likely to be rejected, because the observation of the transit of Mercury on November 4, 1868 by this method had led to very differing results among the various observers, although the progress made in optics over the past century had inspired hopes of greater precision.

Showing remarkable initiative, the committee turned to the heliometer, an excellent model of which was available in all the German observatories. They had been supplied by the workshops of Joseph von Fraunhofer. The heliometer is a telescope, the lens of which is divided into two equal parts, that can move with respect to each other along the separation line. In this way, a double star gives four images, and by superimposing two of them, one can measure the separation of the double system. It was decided that the existing measuring mechanism, a micrometric screw, would be complemented by a second mechanism, reading a graduated scale, thereby making the instrument rather complex and subtle to use.

The advantage of this instrument is to permit the measurement of the distance from the center of the planet and that of the Sun during the whole length of the transit, that is, more than four hours, thus increasing both the probability of observing the phenomenon despite the possible passage of clouds and the precision of the resulting measurement, compared with the contact method.

The eventual recording of the phenomenon by photography was the subject of a very long and very heated debate, including within the photographic sub-committee that had been especially created in order to study the question in depth. Although they were favored by the undeniable success of the photographic observation of the eclipse of the Sun on August 7, 1869, the advocates of this method were a minority for a long time. In order that a final decision be taken, experiments were carried out with variations of the method: dry collodion plates, damp collodion plates, i.e., those generally used for portraits, and daguerreotypes. The reliability of the first technique prevailed over the greater sensitivity of the second, in which the collodion shrinks irregularly while drying. Photography was finally adopted as a second ranked method, which would however be used in all the expeditions, and a very detailed protocol for its use was drawn up, including imprinting a grid on the plates for subsequent measurements.

The contact method was put back on the agenda with the critical examination of the publication of preliminary research work that had been carried out by Mr. Wolf and Mr. André, even though their study was reported in such a succinct way that a sound evaluation could in no case be made. The Parisian astronomers declared with complete assurance that, because of the effects of the spherical aberration, it was impossible to claim precision better than a tenth of a second of arc with this method, unless a lens with a diameter greater than 20 cm was used. Very few German observatories possessed such lenses.

We took advantage of the presence of many astronomers in Leipzig at the end of August, 1872, for the 45th meeting of the Society of German Scientists and Physicians, to organize experiments in measuring contacts with our Fraunhofer 14.5 cm telescopes and a model simulating the planet's transit across the solar disc, supplied by the astronomers of the Imperial Observatory of Pulkovo. Although the appearance of the famous "black ligament" between the two discs just before contact disturbed the measurements, they were satisfactory enough for us to consider the affirmations of the above mentioned gentlemen as presumptuous, and for the contact method to be adopted as the third method of observation, also as a second ranking method.

The sub-committee charged with hiring the personnel of the five expeditions was set up in 1872. Professor Brunhs was a member, and he encouraged his two observers, Carl Börgen and me, to volunteer, and when, after some understandable hesitation, we decided to do so, he supported us strongly. He put forward, with good reason, my proven competence in astronomical photography and in geodesics, and the experience of Börgen who had participated in the German polar expedition in 1869-70; the latter was rapidly named leader of our expedition. It was thus that my distant dream of exercising my profession of astronomer at the other end of the world suddenly became an exciting, and at the same time frightening, reality.

**The trip**

After several days of final preparatory meetings with the executive body, who had come especially to Kiel with two other members of
the committee, we finally left from that port on June 21, 1874,
aboard the Gazelle, a corvette of the Imperial Navy, commanded by Baron von Schleinitz, with officers and a crew of 350 men.

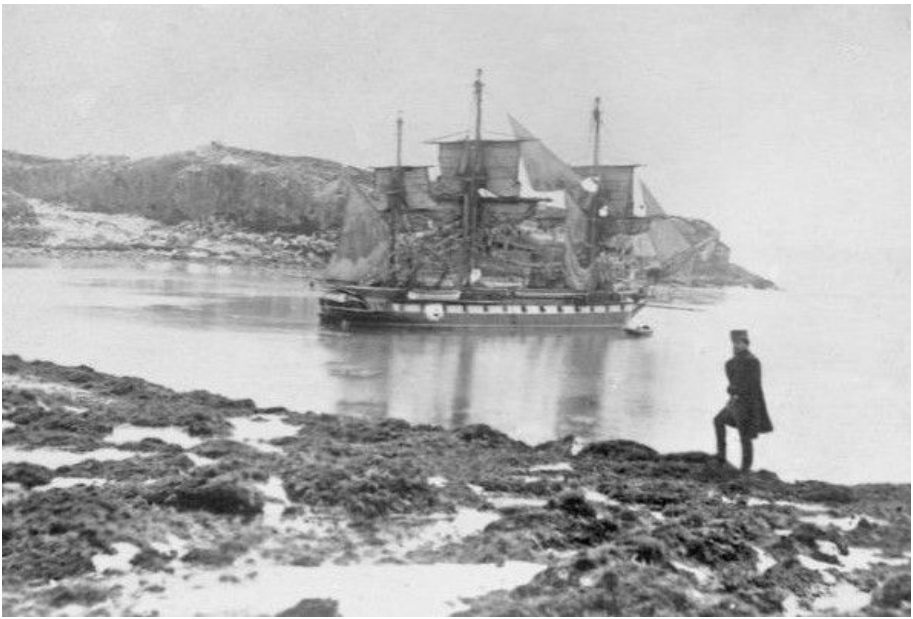
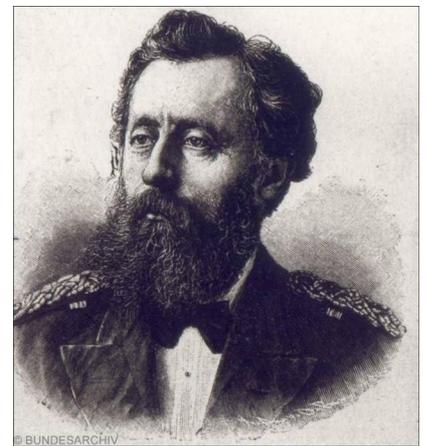

*Baron von Schleinitz[2]*

*The corvette Gazelle anchored at Betsy Cove[3]*

The corvette's mission was to carry the six members of our expedition, with all their instruments, as well as the parts of two iron observation towers and a wooden house, to the Kerguelen Islands, carrying out soundings and hydrographic measurements on the way. After the completion of our observations, it was to carry us back to the Island of Mauritius, from which we were to find our own way back to Europe, while it carried out its own scientific mission around the world until April, 1876. This was the first mission of this kind for the young Imperial Navy.

For the 26-year-old that I was at the time, who had never left his hometown, Budapest, except to study in Vienna and in Northern Germany, the trip, which lasted more than three months, with stops in Madeira, the Cape Verde Islands, Monrovia, Ascension Island, the mouth of the Congo River, Capetown, was absolutely unforgettable. The new unimaginable landscapes, the heat, the animals, the fish, the exotic birds, the strange fruit, of which our zoologist, who had been delegated as our doctor, rapidly gave us the Latin name, and sometimes the popular name, the Negroes with their strange customs - everything was a source of almost daily amazement, which I attempted to cristallize by numerous sketches and drawings. The passage of the Moon at the zenith, the appearance of unknown constellations on the firmament, the apparent motion of the Sun and the stars in a direction opposite to that to which astronomers are accustomed, accentuated my feeling of strangeness. The repeated storms that we weathered beyond the Cape of Good Hope, and the death of a sailor who smashed his head on the bridge falling from the rigging during one of them, caused me strong and lasting emotions, such as I had never felt before.

---

2   © 2020 Bundesarchiv, Bild 134/B 2957
3   © 2020 Bundesarchiv, Bild 134/C161

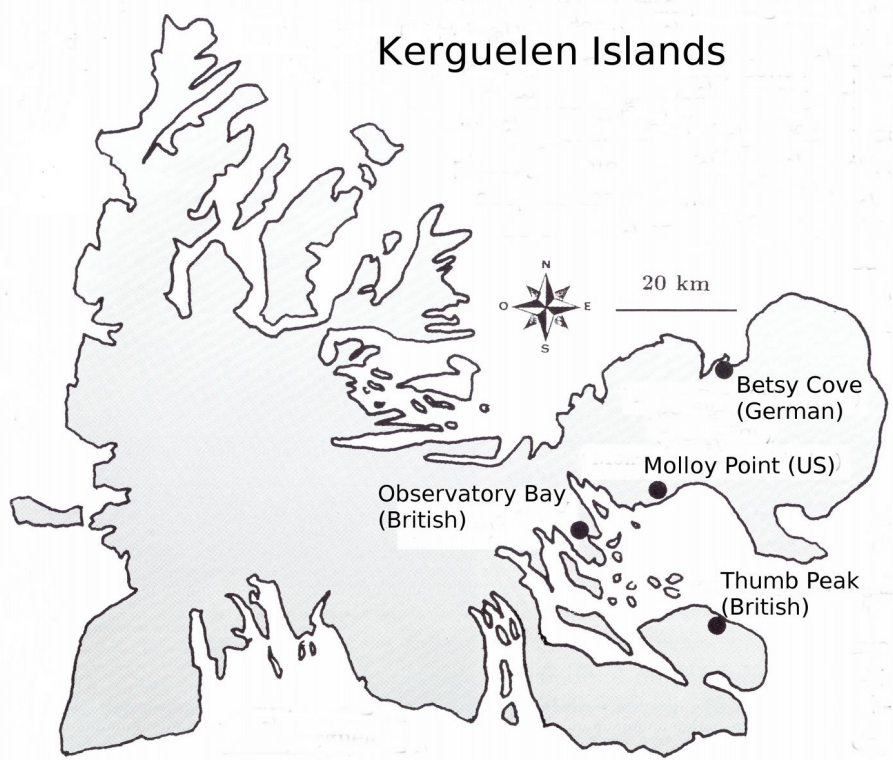

**The Kerguelen Islands**

After spending several days waiting for a lull, we landed at Betsy Cove, a small inlet of Accessible Bay on the northeastern coast of the big island, on October 26, 1874, a beautiful sunny morning, only 102 years after the island was discovered by Captain de Kerguelen de Trimarec.

Once we had chosen a suitable location, sheltered from the predominating winds, close to the shore and to a spring of sweet water, with good visibility in the direction of the phenomenon, and once we had smoothed out the ground as well as we could, the equipment was transported there, with the heavy crates dragged along makeshift wooden rails, the buildings were set up by the good sailors, who were stopped neither by rain, nor snow, nor icy winds, and God knows that there were all of these.

Other than bad weather, we did not have any particular difficulties. This was not the case for the German expedition in China, who had misplaced the assembling instructions! In two weeks, the German flag flew above our new lodgings, which we inaugurated with music, finally leaving our rocking quarters aboard the Gazelle for the windier ones on the ground.

**Preparing for the observations**

My first task was to align the instruments with the meridian, which I did starting on October 29th,

by measuring the height of the Sun at moon and toward sunset with the help of a prism circle and by carefully writing down the time of the chronometers, not without several interruptions while I took shelter from the sudden flurries of snow.

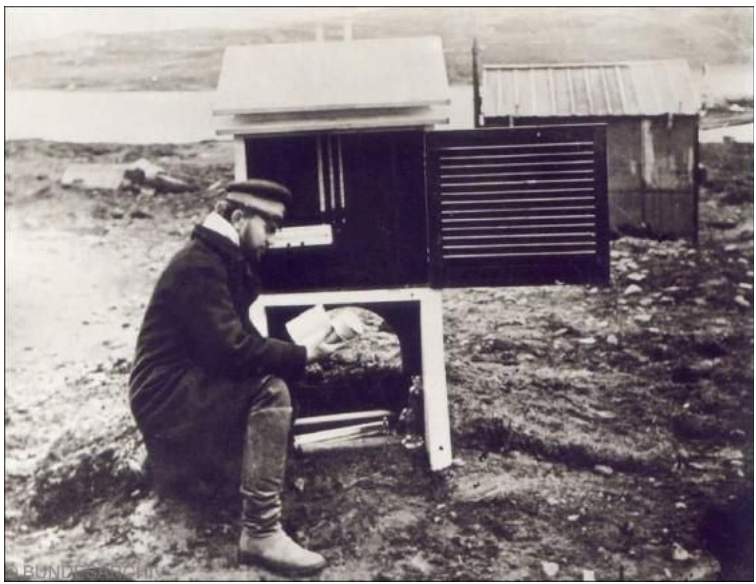

*Lieutenant Wachenhusen examining the meteorological instruments[4]*

The following weeks were spent installing the three large observing instruments and aligning their axes with that of the Earth: a heliometer with a 9.5 centimeter opening, a Fraunhoper refractor with a 12 centimeter opening to observe the contacts, and a refractor with a 16 centimeter opening for the photographs. They were fragile instruments, all of them being at least fifty years old, and cumbersome, with their 2 or 3 meter long tubes.

On November 16, for the first time I observed the polar star σ-Octantis, and obtained the first time measurement at the transit instrument. On November 22, with a full moon, a storm foiled my attempts to measure the Moon's culmination. The wind shook the instrument, blew out the lamps, and I could not even hear the beats of the chronometer near my ear.

Under these persistent meteorological conditions, astronomical observations to check the time of our chronometers and to determine the coordinates of the situation were a particularly frustrating activity. Most nights, I spent five or six hours at the telescope waiting for a hole in the clouds and a lull in the blasts of wind, at the same time, without ever really obtaining satisfactory results. It sometimes even happened that, with the storm blowing out of all proportion, we had to rush out of our house in order to solidly attach the two iron observation towers with ropes, and to place on the ground all the glass containers in the dark room. Fortunately, it was not too cold, the temperature always remaining at about 4 degrees. But this was nevertheless high summer!

The frequency of the storms, on an average every other day, was not a good omen for the day of the phenomenon, and we could not help being worried. The otherwise very interesting visit by Captain Fuller on December 3, who had been seal hunting in the area for sixteen years, did not reassure us, because he mentioned that bad weather was common when the Moon changes. The new Moon was due exactly on the 9th.

Preparations intensified during the last week. Exercises in measuring contacts with the set-up simulating the transit of Venus, training in automatically measuring and in reading the scale on the heliometer. The worst was certainly the very careful preparation of a stock of about a hundred dry collodion plates, at about ten per hour. ln complete darkness, one observer dusted the plates, spread them with collodion, put silver on them, then soaked them in a bath of sterilized water, while another observer shook them, then after carefully rinsing them, covered them with a solution of ammoniated albumin, silver and nitric acid, after which the plates were left to dry for three or four hours. To make the albumin, we used eggs from the hens we had brought from Capetown, after giving up on dry albumin and on that from penguin eggs.

---

4   © 2020 Bundesarchiv, Bild 134/

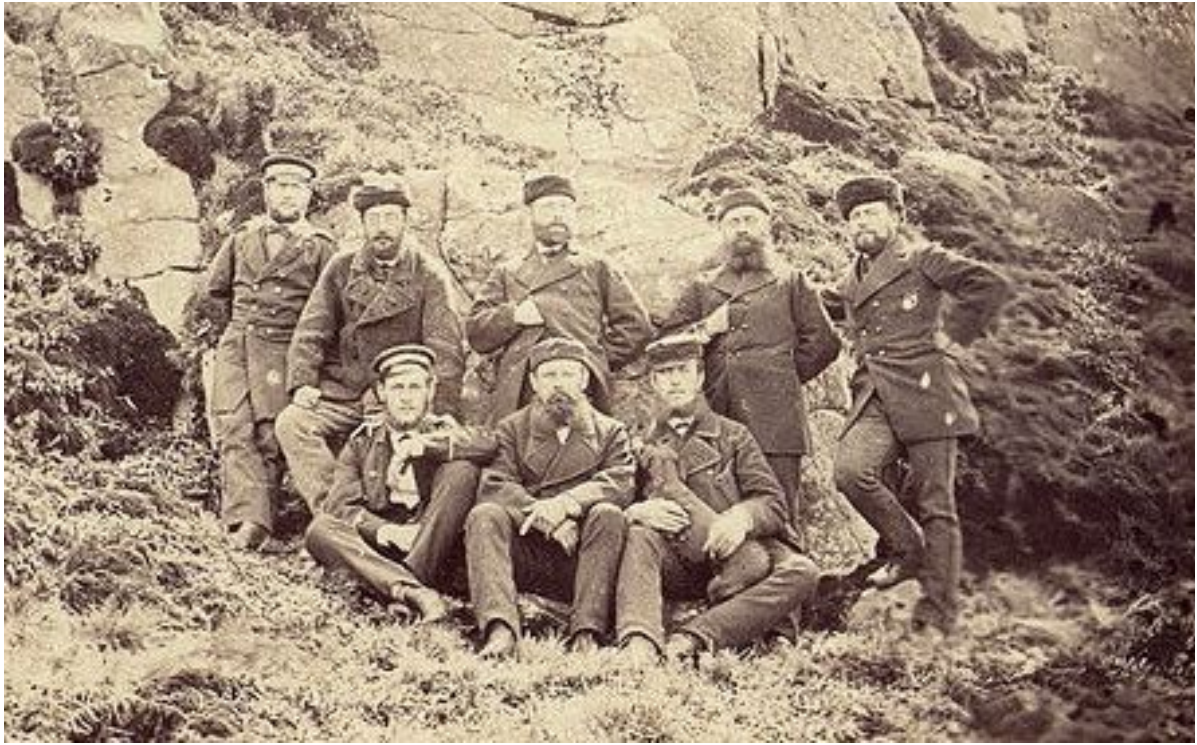

*The crew of the Gazelle on the Kerguelen islands*[5]

**December 9, 1874**

On the 7th there was a storm and on the 8th, it was still raining, but on the 9th, a friendly Sun rose in a clear sky and a calm atmosphere. Venus's first contact with the solar disc was to happen toward half past six in the morning. A few minutes before the anticipated moment, we were all in position: Bôrgen at the heliometer, I at the Fraunhofer refractor with the mechanic, Krille, who was to follow the Sun, Wittstein at the refractor for the contact measurements, and Lieutenant von Ahlefeld at the small 21 centimeter terrestrial telescope, along with a sailor at each telescope in charge of lighting. During this time, the sky began to be covered with cirrus and cirrostratus clouds, which made us miss the first contact but did not destroy our hopes, because we soon saw the little black disc of the planet progressively penetrate the luminous disc of the Sun. With heightened tension we waited for the contact of the second edge of the planet with that of the Sun.

The critical moment approached, everyone looking at his chronometer again in order to control his mental counting of the seconds. Venus seemed to want to detach itself from the edge of the Sun, a bridge formed, became thinner, and suddenly disappeared. This was the moment we had all been waiting for; we had finally managed to get it. While Venus was transiting the solar disc, the astronomers' tasks changed. Börgen stayed at the heliometer, now helped by Wittstein, who read the graduated scale. I was at the photographic refractor, still aided by Krille, who changed the plate holders and carried them into the darkroom, where two assistants prepared the plates and loaded them into the plate holders.

---

5 © Dr. Jens Mattow (Image Luminous Lint/88631)

During the four and a half hours that the transit lasted, we exposed 21 damp plates and 40 dry ones. We would have taken twice as many, had not ever thicker clouds considerably dimmed the Sun's brightness during the second half of the transit. Despite this, we managed to correctly observe the last two contacts. We were also lucky to have an exceptionally calm atmosphere; the high wind which had been the rule throughout our stay would have strongly disturbed the instruments and called into question the precision of the measurements.

Once the phenomenon was over, I could not help admiring the extent of human knowledge, that had allowed us to predict, within a few minutes, an event observed for the last time a hundred and five and a half years earlier. l admired at the same time the trust of our governments, who had agreed to finance very costly expeditions, and that of astronomers who, after a long and sometimes difficult journey, arrived in a desolate place where nothing indicates the possible occurrence of the phenomenon, yet nevertheless prepared to observe it. A few days earlier, Venus, the Splendid evening star, had shone on the western horizon; but, on the day and the time predicted, it transited across the Sun!

**Epilogue**

The expedition left the large island for good on February 5, 1875, after spending another two months calibrating the scale of the heliometer by measurements of the Sun's diameter, and determining the precise coordinates of the observatory by polar and zenith distances, culminations of the Moon and occultations of stars.

The results of the campaigns of 1874 and 1882 were very disappointing because they were not much better than those of the preceding century, probably because of unforeseeable effects of the refraction of solar rays in the planet's dense atmosphere. The distance from the Earth to the Sun is now known with very great accuracy, thanks to radar telemetry, and the next transits of Venus across the Sun, on June 7, 2004 and June 5, 2012, will certainly take place amid general indifference, unless astronomers decide to celebrate this phenomenon, which held such an important place in their history.

**Acknowledgments.** This is the translation of a paper in French published in *Pulsar*, n°733, p.8-12, 1999. I thank Barbara Jachowicz for the translation.